\documentclass{mn2e}
\usepackage{natbib}
\usepackage{amsfonts}
\usepackage{amsmath}
\usepackage{amssymb}
\usepackage{graphicx}

\newcommand{\del}{\ensuremath{\delta}}
\newcommand{\Del}{\ensuremath{\Delta}}
\newcommand{\lam}{\ensuremath{\lambda}}
\newcommand{\Lam}{\ensuremath{\Lambda}}
\newcommand{\gam}{\ensuremath{\gamma}}
\newcommand{\Gam}{\ensuremath{\Gamma}}
\newcommand{\sig}{\ensuremath{\sigma}}

\newcommand{\epc}{\ensuremath{\epsilon_{\times}}}
\newcommand{\Sc}{\ensuremath{S_{\times}}}
\newcommand{\So}{\ensuremath{S_{0}}}
\newcommand{\delc}{\ensuremath{\delta_{\rm c}}}
\newcommand{\delo}{\ensuremath{\delta_{0}}}

\newcommand{\rhoh}{\ensuremath{\rho_{\rm h}}}
\newcommand{\dprbar}{\ensuremath{{\bar\delta^\prime}}}
\newcommand{\sigdpr}{\ensuremath{\bar\sigma}}

\newcommand{\sigdprsq}{\ensuremath{\bar\sigma^2}}
\newcommand{\dcr}{\ensuremath{\delta_{{\rm c}\times}}}

\newcommand{\avg}[1]{\ensuremath{\left\langle \,#1\, \right\rangle}}

\newcommand{\der}{\ensuremath{{\rm d}}}

\newcommand{\cb}{\ensuremath{\mathbf{c}}}

\newcommand{\erf}[1]{\ensuremath{{\rm erf}\left(#1\right)}}

\newcommand{\eqn}[1]{equation~\eqref{#1}}
\newcommand{\eqns}[1]{equations~\eqref{#1}}
\newcommand{\fig}[1]{Figure~\ref{#1}}
\newcommand{\figs}[1]{Figures~\ref{#1}}
\newcommand{\ph}[1]{\phantom{#1}}

\newcommand{\be}{\begin{equation}}
\newcommand{\ee}{\end{equation}}
\newcommand{\Cal}[1]{\ensuremath{\mathcal{#1}}}

\title[Excursion set peaks]
      {Peaks theory and the excursion set approach} 

\author[A. Paranjape \& R. K. Sheth]
{Aseem Paranjape$^{1}$\thanks{E-mail: aparanja@ictp.it} \& Ravi K. Sheth$^{1,2}$\\  
 $^1$ The Abdus Salam International Center for Theoretical Physics,
      Strada Costiera, 11, Trieste 34151, Italy\\
 $^2$ Center for Particle Cosmology, University of Pennsylvania, 
      209 S. 33rd St., Philadelphia, PA 19104, USA}
\begin{document}
\pagerange{\pageref{firstpage}--\pageref{lastpage}}

\maketitle 

\label{firstpage}

\begin{abstract}
\noindent 
We describe a model of dark matter halo abundances and clustering 
which combines the two most widely used approaches to this problem:  
that based on peaks and the other based on excursion sets.  
Our approach can be thought of as addressing the cloud-in-cloud 
problem for peaks and/or modifying the excursion set approach so that 
it averages over a special subset, rather than all possible walks.  
In this respect, it seeks to account for correlations between steps 
in the walk as well as correlations between walks.  
We first show how the excursion set and peaks models can be written 
in the same formalism, and then use this correspondence to write our 
combined excursion set peaks model.  
We then give simple expressions for the mass function and bias, 
showing that even the linear halo bias factor is predicted to be 
$k$-dependent as a consequence of the nonlocality associated with 
the peak constraint.
At large masses, our model has little or no need to rescale the 
variable $\delc$ from the value associated with spherical collapse, 
and suggests a simple explanation for why the linear halo bias 
factor appears to lie above that based on the peak-background split 
at high masses when such a rescaling is assumed.  Although we have 
concentrated on peaks, our analysis is more generally applicable to 
other traditionally single-scale analyses of large-scale structure.  
\end{abstract}

\begin{keywords}
large-scale structure of Universe
\end{keywords}

\section{Introduction}
\label{intro}

\citet{ps74} argued that the abundance of nonlinear virialized objects at late times (such as the present) should be sensitive to the statistics of the initial fluctuation field, and to the subsequent expansion history of the universe.  This is the basis for studies which seek to use the abundance and clustering of galaxy clusters to constrain cosmological parameters.  

Their work has motivated the study of analytical models for the formation, and hence the abundance and spatial distribution, of halos, which can be used to provide fitting formulae when interpreting data.  
Following \cite{st99}, the most widely used fitting formulae are self-similar, in the sense that the predicted halo abundances can be scaled to a universal form which is independent of cosmology, redshift and power spectrum.  This vastly simplifies cosmological analyses.  (This universality is only expected to hold approximately, and the next generation of datasets may have sufficiently many clusters that departures from universality must be accounted for.  We will have more to say about this later.)

The self-similar functional form can be derived from a physically motivated model of collapse \citep*{smt01}. The number density $\der n/\der m$ of halos in the mass range $(m,m+\der m)$ is written as
\be
 \frac{m}{\bar\rho}\frac{{\rm d}n(m)}{{\rm d}m} {\rm d}m = f(\nu)\,\der\nu, 
 \label{excsetansatz}
\ee
where $\bar\rho$ is the background density and $\nu=\delta_c/\sigma$, with $\delc$ the rescaled time variable and $\sigma$ the rescaled mass variable ($\sigma^2(m) \equiv \avg{\del^2(m)}$ is the variance of the matter density field smoothed on a Lagrangian length scale corresponding to mass $m$ and linearly extrapolated to present day). Universality is manifest in the statement that $f$ depends only on $\nu$, but, unfortunately, the most naive use of this form predicts too few massive clusters.  This has motivated the following ad-hoc approach: one actually fits $f(\sqrt{q}\nu)$ to the data, and determines $q$ from the fit.  This semi-empirical approach has worked rather well, in the sense that $q$ appears to be approximately independent of cosmology, redshift and power spectrum, although recent simulations are beginning to show departures from universality \citep{bkk09}.  

Since observations will soon deliver large cluster catalogs over a range of redshifts, it is clearly desirable to have a more fundamental understanding of why $q\ne 1$, particularly because, on an object by object basis, the physical model of collapse almost never has $\delta<\delc$.  I.e., $q<1$ appears to arise in the step which converts from the physics of halo formation to a statistical description of halo abundances \citep{smt01}.  One of the main goals of this paper is to provide some insight into the origin of this factor.  

To do so, we revisit the two most common models for identifying halos from the initial fluctuation field:  the peaks theory of \citet[hereafter BBKS]{bbks86}, and the excursion set approach of \citet{bcek91}.  Although both make predictions which can be phrased in terms of the self-similar variable $\nu$, and both treat $\nu$ as the ratio of $\delta_c/\sigma$, the former treats the numerator of this ratio as the stochastic quantity, whereas for the latter, it is the denominator which can vary.  They also differ fundamentally in their approach to the problem.  Peaks theory seeks to describe the point process which describes the special positions in the initial conditions around which halos collapse.  The excursion set approach aims only at a statistical description of the mass fraction in bound objects, and assumes that this can be done by consideration of all points in space -- not just the special ones around which halos form.  Our analysis below shows how to merge the two descriptions.  

Section~\ref{mf-uncond} shows that the excursion set and peaks models can be written in the same formalism, and then describes our excursion set model for peaks, arguing that the result goes a substantial way towards explaining the origin of the factor of $a$.  Section~\ref{mf-cond} extends this to describe the conditional function of excursion set peaks in constrained larger-scale environments, and from there builds a model for the large scale bias factors.  This uses the recent work of \cite*[hereafter MPS]{mps12} to show that halo bias in our approach is generically expected to be $k$ dependent.  It also shows that at high masses, our new expression for halo bias is qualitatively similar to that seen in simulations, again suggesting that our excursion set peaks model of the origin of $a\ne 1$ is reasonable.  A final section summarizes our results, discusses them in the context of previous work on the relationship between excursion sets and peaks, and suggests ways in which our approach could be improved further.

\section{The unconditional mass function}
\label{mf-uncond}

This section develops what we call the excursion set model for peaks:  it is both a peaks model which deals with different smoothing scales, and an excursion set model which deals with the statistics of special rather than random positions.  This is particularly interesting because \cite{smt01} have argued that the latter is a necessary change to the standard excursion set approach, and \cite{lp11} have argued that the correspondence between peaks in the initial conditions and halos at late times in their simulations is quite good.  

\subsection{Notation}
\label{notation}
Let $s(R)$ denote the variance of the (linearly extrapolated) density 
contrast \del\ smoothed on a Lagrangian length scale $R$.
Then $s\equiv \sigma_0^2$ where 
\be
 \sig_j^2 = \int \frac{\der^3k}{(2\pi)^3}\,P(k)\,k^{2j}\,W^2(kR)
\ee 
and our notation has dropped the explicit dependence of $\sigma_j$ on 
$R$, where we think no confusion will arise.  Here $P(k)$ is the 
power spectrum of the field, and $W(kR)$ is the Fourier transform of 
the smoothing filter.  For what follows, it is also convenient to 
define 
\be
 R_\ast \equiv \sqrt{3}\sig_1/\sig_2 \qquad {\rm and}\qquad  
 \gam \equiv \sig_1^2/\sig_0\sig_2\,.
\label{VstRstgam}
\ee
Unless stated otherwise, we will always consider Gaussian smoothing 
filters, for which $W(kR)=\exp(-k^2R^2/2)$, and, for the $P(k)$ of current 
interest in cosmology, all the integrals above converge.  

We return to the issue of smoothing filter in the Discussion section, 
because the analysis which follows exploits the following property which 
is special to Gaussian smoothing.  Namely, the Laplacian of the field 
on a given smoothing scale $\nabla^2_{\bf x} \delta(R,{\bf x})$, 
which is a quantity of fundamental importance in peaks theory, is 
the same as the derivative of the field with respect to smoothing 
scale, a fundamental quantity in excursion set theory.  

On dimensional grounds, the volume associated with a smoothing filter 
is $V\propto R^3$; for a Gaussian filter $V = (2\pi R^2)^{3/2}$.  
It is natural to associate a mass with the smoothing scale $R$:  
$m\equiv \bar\rho V$, where $\bar\rho$ is the comoving background 
density.  
We will only consider hierarchical models in which the fluctuations 
in the initial field were small.  The former means that $\sigma_0$ 
is a monotonically decreasing function of $R$, and the latter that 
$R$, $m$ and $\sigma_0$ are equivalent variables.  
In what follows we illustrate our results using $P(k)$ for a 
flat \Lam CDM cosmological model with parameters 
$(\Omega_m,\Omega_\Lam,h,\sig_8,n_s) = (0.25,0.75,0.7,0.8,0.95)$.

\subsection{The excursion set approach}
The excursion set approach assumes that the mass fraction associated with bound halos of mass $m$ at any given time $t$ equals the volume fraction of positions in the initial fluctuation field which, when smoothed on scale $R\propto m^{1/3}$, have overdensity $\delta(R)=\delc(t)$ and, for all $R'>R$, have $\delta(R')< \delc(t)$.  This latter constraint is difficult to handle because it implies an infinite number of constraints (one for each smoothing scale).  

\citet[hereafter MS]{ms12} showed that the much simpler requirements that $\delta(R)>\delc$ and $\delta(R + \Del R)<\delc$ for $\Del R\ll 1$ (i.e., just one additional constraint) permit a simple analytic estimate of this fraction which is remarkably accurate.  Although this fraction depends on the power spectrum of the underlying fluctuation field, much of this dependence can be removed if one works instead with the requirement $\delta(s)>\delc$ and $\delta(s-\Del s)<\delc$ (recall that $s$ and $R$ are equivalent variables).  Thus, if $v\equiv \der\delta/\der s$, then MS argued that the fraction $f(s)$ of interest satisfies 
\be
 \Del s\,f(s)\equiv \int_0^\infty \der v\,
                    \int_{\delc}^{\delc + v\Del s} \der\delta\,p(v,\delta) \,,
\ee
where $p(v,\del)$ is the joint distribution of $\delta$ and its derivative $v\equiv\der\del/\der s$.  Taking the limit $\Del s\to \der s\ll 1$ implies 
\be
 f(s)\equiv \int_0^\infty \der v\,v\,p(v,\delc) \,,
 \label{fMS}
\ee
%
%This follows from writing the probability that the height $\del(s)$ lie between \delc\ and $\delc+ v\Del s$ with $v>0$, and interpreting the limit $\Del s\to \der s$ as the required fraction $\Del s\, f(s)\to \der s\, f(s)$. 
We will see shortly that the issue is what exactly to use for $p(v,\del)$.  

The rms values of $\delta$ and $v$ are $\sigma_0$ and $(2\gamma\sigma_0)^{-1}$, respectively.  So, if we define 
\be
 \nu\equiv \delc/\sigma_0\qquad {\rm and}\qquad x\equiv 2\gamma\sigma_0\,v,
 \label{xnu}
\ee
(our choice of notation will become clear shortly) and if we choose to average over all positions in the initial Gaussian random field which have height $\delc$ on scale $s$, then equation~(\ref{fMS}) implies that 
\be
 sf(s)\equiv \frac{\exp(-\nu^2/2)}{2\gamma\sqrt{2\pi}}\,
            \int_0^\infty \der x\,x\,p_G(x-\gamma\nu;1-\gamma^2) 
 \label{sfs-ms}
\ee
where $p_{\rm G}(y-\mu;s)$ a Gaussian distribution for variable $y$, 
with mean $\mu$ and variance $s$.  The right hand side is clearly a 
function of the scaling variable $\nu$ and $\gamma$.  The dependence 
on $\gamma$ means that the result is not a completely universal function 
of $\nu$, but MS argued that, over the range of power spectra of current 
interest in cosmology, this dependence is relatively weak.  
Therefore, it is useful to use the fact that 
 $\nu f(\nu) = s f(s)\, |\der\ln s/\der\ln\nu| = 2s f(s)$ 
to write the expression above in terms of the scaling variable $\nu$:
\be
 \nu f_{\rm MS}(\nu)\equiv \frac{\nu\,\exp(-\nu^2/2)}{\sqrt{2\pi}}\,
              \frac{\avg{x|\gam,\gam\nu}_{\rm MS}}{\gam\nu}\,, 
 \label{vfv-ms}
\ee
where
\be 
 \avg{x|\gam,x_\ast}_{\rm MS}\equiv 
                   \int_0^\infty \der x\,x\,p_G(x-x_\ast;1-\gamma^2).
 \label{avgx-ms}
\ee
This shows that $f(\nu)$ is the product of the Gaussian probability 
of having height $\nu$ and the mean `curvature' associated with such 
positions.  (This integral can be done analytically; see MS.  
Also, although we will not focus on the expected departures from 
universality in this model, which come from the dependence on $\gamma$, 
we would like to emphasize that small departures are predicted.)  

MS showed that this expression, based on their `one-step' 
approximation, provided an excellent description of the exact 
solution in which the constraint on the walk height is satisfied 
on all scales.  While this is significant -- it effectively solves 
the same excursion set problem for `correlated' steps that \cite{bcek91} 
solved for `uncorrelated' steps -- it does not really solve the 
problem as stated at the beginning of this sub-section.  Namely, 
the quantity of interest is a volume fraction in the initial field.  
One should estimate this by averaging over the full set of walks in 
one realization of the field.  But what has actually been 
calculated is an ergodic average over an ensemble of walks in 
which the steps in each walk are correlated, but the walks themselves 
are independent of one another.  
\cite{smt01} demonstrated that this replacement is incorrect; 
in fact, one must either account for the correlations between walks, 
or account for the fact that the set of walks over which one should 
average is a special subset of all walks \cite*[also see][]{rks11,pls12}.  

\begin{figure}
 \centering
 \includegraphics[width=\hsize]{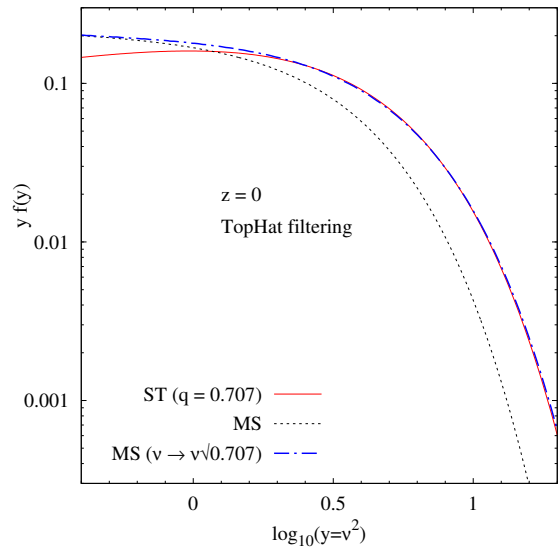}
 \caption{Mass function of halos identified in a \Lam CDM simulation 
   \citep[solid red;][with their $q=0.707$]{st99} and as predicted 
   by the excursion set approach \citep{ms12} 
   (dotted black) and with $\nu\to\sqrt{0.707}\nu$ (dot-dashed blue) 
   in \eqn{vfv-ms}. 
   This plot uses TopHat filtering of the \Lam CDM power spectrum.
   }
 \label{vfv-th}
\end{figure}

Figure~\ref{vfv-th} provides one illustration of why the distinction 
matters.  (For this plot only, we have used a Tophat, rather than Gaussian, 
filter in all integrals over $P(k)$.)  The solid curve shows the 
fitting function of \cite{st99} 
\be
 \nu\,f_{\rm ST}(\nu) = 0.644\,\Bigl[1 + (q\nu^2)^{-0.3}\Bigr]\,
  \frac{\sqrt{q\nu^2}\,\exp(-q\nu^2/2)}{\sqrt{2\pi}}
 \label{fST}
\ee
which provides a good description of halo counts in simulations.  
As mentioned in the Introduction, this 
function is expressed in terms of the scaling variable $\sqrt{q}\nu$, 
where $\nu$ is the same quantity which enters in the excursion set 
approach and $q\approx 0.7$.  The dotted curve shows \eqn{vfv-ms}; it 
vastly underestimates the halo counts at large $\nu$.  

It is easy to see why this happens.  At $\nu^2\gg 1$, 
%the halo counts become 
 $\nu\,f_{\rm ST}(\nu)\to 0.644 \sqrt{q\nu^2}\,\exp(-q\nu^2/2)\sqrt{2\pi}$ 
whereas \eqn{vfv-ms} becomes 
 $\nu\,f_{\rm MS}(\nu)\to \nu\,\exp(-\nu^2/2)/\sqrt{2\pi}$.  
This shows that if we rescale $\nu\to \sqrt{q}\nu$ in \eqn{vfv-ms}, 
then the result should provide a good description of the halo counts, 
upto the difference in amplitude of 0.644.  However, even this difference 
in amplitude can be accounted for by noting that the term in square 
brackets in \eqn{fST} only equals unity at very large $\nu$.  
At $\nu^2=10$, it is $\approx 3/2$, so multiplying by 0.644 yields 
unity.  The dot-dashed curve shows that, indeed, rescaling 
$\nu\to \sqrt{0.707}\nu$ in \eqn{vfv-ms} works very well.  

Note that $q<1$ is required to achieve this agreement at large $\nu$.  
While it is tempting to associate this rescaling with a reduction in 
the value of \delc, this is problematic because direct measurements 
of the overdensity in patches which are destined to form halos show 
that the critical density required for collapse increases at small 
$\nu$.  This increase is qualitatively consistent with expectations 
based on modelling halo formation using a triaxial rather than 
spherical collapse \citep{smt01}.  I.e., the physics of halo formation 
suggests that, if we want to think of the parameter $q$ as rescaling 
$\delc$, then $q$ should be greater, rather than less than unity.  
The alternative, which we will not explore further here \cite[but see 
discussion in][]{pls12}, is to assume that $\sqrt{q}$ rescales $\sigma$ 
rather than $\delc$.  Rather, the next 
section explores a quite different reason for why rescaling with $q<1$ 
works so well.  The main point we wish to make here is that, absent an 
understanding of why rescaling $\nu$ was necessary, one should not 
use the shape of $f(\nu)$ to make conclusions about whether the 
physics of collapse was spherical or not; the evidence for triaxial 
collapse comes from the direct measurements of the properties of the 
patches from which halos formed \citep[i.e., those in][]{smt01}.  

\subsection{Peaks in the initial field}
In the argument which led to \eqn{vfv-ms}, we noted that the predicted 
mass fraction $f(\nu)$ depends critically on what one chooses for 
$p(v|\delc)$.  In what follows, we will show what happens if we wish 
to add the additional constraint that, on scale $s$, 
$\delta = \delc$ is also a local maximum of the field.  

The number density of peaks of height $\delta$ depends critically on 
the smoothing scale on which the peaks were defined.  If $\sigma_0$ 
denotes the rms value of the density fluctuation on the chosen 
smoothing scale, then the number density of peaks of scaled height 
$\nu=\del/\sigma_0$ in a Gaussian-smoothed Gaussian random field is 
\be
 \Cal{N}_{\rm pk}(\nu) = \int \der x\, \Cal{N}_{\rm pk}(x,\nu)
      = \frac{{\rm e}^{-\nu^2/2}}{\sqrt{2\pi}}\,
        \frac{G_0(\gam,\gam\nu)}{(2\pi R_\ast^2)^{3/2}} ,
\label{Npk-bbks}
\ee
where $\gamma$ and $R_\ast$ were defined earlier, and 
\be
 G_J(\gam,x_\ast) \equiv \int_0^\infty\der x\, x^J
                         F(x)\,p_{\rm G}(x-x_\ast;1-\gam^2)\,, 
\label{Gbbks}
\ee
with
\begin{align}
F(x)&=\frac12\left(x^3-3x\right)\left\{\erf{x\sqrt{\frac52}}+\erf{x\sqrt{\frac58}}
  \right\} \notag\\
&\ph{x^3-3x}
+ \sqrt{\frac2{5\pi}}\bigg[\left(\frac{31x^2}{4}+\frac85\right){\rm
    e}^{-5x^2/8} \notag\\
&\ph{\sqrt{x^3-3x+\frac2{5\pi}}[]}
+ \left(\frac{x^2}{2}-\frac85\right){\rm
    e}^{-5x^2/2}\bigg]\,,
\label{eqn-bbks-Fx}
\end{align}
(equations~A14--A19 in BBKS).
The variable $x$ is the Laplacian of the field normalized by its 
rms value (for Gaussian filters, this rms is $\sigma_2$), so it 
represents the curvature around the peak position.  Therefore, 
$F(x)$ quantifies how different the set of curvatures is around a 
peak position compared to a randomly placed one.  

To map from peak number densities to halo mass fractions, one must 
associate a mass with each peak.  The natural choice is the mass $m$ 
contained within the smoothing window:
  $m = \bar\rho\,V$ with $V\propto R^3$.  
But this has the unfortunate consequence that peaks of different 
height $\nu$ will all have the same mass, whereas the intuitive 
expectation is that more massive objects should be associated with 
higher peaks.  

I.e., the intuitive picture is one in which there is a critical 
density contrast $\delc$ which is associated with halos (in what 
follows we will set $\delc=1.686$, thus ignoring the mild dependence 
on cosmology predicted by the spherical collapse model), and massive 
halos have large $\nu = \delc/\sigma_0$ because they have small 
$\sigma_0$ (this is what happens naturally in the excursion set approach).  
Thus, the main difficulty in identifying peaks with halos is that 
\eqn{Npk-bbks} is defined for a fixed smoothing scale $R$ (so 
changes in $\nu$ are due to changes in $\delta$), whereas one would 
really like to allow $R$ to vary instead.  

If one assumes naively (and incorrectly) that $\nu$ in 
\eqn{Npk-bbks} has $\delc$ fixed and $R$ varying, then one might 
naively (and incorrectly) assume that the mass fraction of the 
Universe that is in peaks of mass $m$ is given by 
\be
 f_{\rm BBKS}(\nu) = \frac{m}{\bar\rho}\,\Cal{N}_{\rm pk}(\nu) = 
\frac{{\rm e}^{-\nu^2/2}}{\sqrt{2\pi}}\frac{V}{V_\ast}\, G_0(\gam,\gam\nu)\,,
\label{vfv-bbks}
\ee
where we have defined $V_\ast = \left(2\pi R_\ast^2 \right)^{3/2}$.
Quite apart from the mathematical inconsistency associated with 
making this assumption, there is a conceptual difficulty which is 
known as the cloud-in-cloud problem.  This comes from considering 
how the density around a given position fluctuates as one changes 
the smoothing scale $R$.  One might imagine that a given position 
is a peak on some smoothing scales and not on others; or that a 
position which is a local maximum of height $\nu$ on a small 
smoothing scale may have an even larger value of $\nu$ on a large 
smoothing scale, without being a local maximum of the field on the 
larger smoothing scale.  Which, if any of these cases, should one 
associate with halos?  

\subsection{Excursion set peaks}
As MS noted, the excursion set approach of the previous section shows 
how one might address this problem more consistently.  
Namely, it says that of the peaks present on scale $s$, we want 
those which have a smaller height on the next larger smoothing scale.  
Therefore, the same logic which led to \eqn{sfs-ms} (i.e., demanding that the scaled peak height lie between $\nu=\delc/\sig_0$ and $\nu+(x/2\gam)\Del\ln s$) will now yield
\be
\Cal{N}_{\rm ESP}(\nu) =
\frac{1}{\gam\nu}\int_0^\infty\der x\,x\,\Cal{N}_{\rm pk}(\nu,x)\,,
\label{Nesp}
\ee
making 
\be
 f_{\rm ESP}(\nu) = \frac{{\rm e}^{-\nu^2/2}}{\sqrt{2\pi}}
                          \frac{V}{V_\ast}\, G_0(\gam,\gam\nu)
                   \frac{\avg{x|\gam,\gam\nu}_{\rm ESP}}{\gam\nu}\,, 
\label{vfv-esp}
\ee
where
\be
 \avg{x|\gam,x_\ast}_{\rm ESP} = G_1(\gam,x_\ast)/G_0(\gam,x_\ast).
\label{avgx-esp}
\ee
(This is where the choice of a Gaussian for the smoothing filter 
simplifies the analysis, since a constraint on the value of the 
derivative with respect to smoothing scale becomes a constraint on 
the curvature of the field.  This also explains why, in \eqn{xnu} 
we used $x$ to denote $\der \del/\der s$ normalized by its rms value.)

Essentially this same formula for peaks, \eqn{vfv-esp}, has appeared 
previously \citep{aj90}.  However, we believe our treatment 
highlights the similarities and differences between peaks and 
random positions more clearly. In particular, notice that 
$f_{\rm ESP}$ modifies the peaks probability in the same way that 
$f_{\rm MS}$ modifies the gaussian pdf:
the distribution of peaks picks up an additional factor of the 
normalised mean \emph{peak} curvature.  In this respect, the only 
conceptual difference between $f_{\rm ESP}$ and $f_{\rm MS}$ is that 
the latter averages over `random' positions in the field, whereas the 
former averages over `special' ones.  In other words, $f_{\rm ESP}$ 
addresses both the cloud-in-cloud problem for peaks (the fundamental 
failing of the peaks approach), 
and the question of how the excursion set predictions are modified 
if one averaged over special positions in the initial field (the 
fundamental failing of the excursion set approach).  

The analysis above shows that changing the ensemble over which 
the excursion set average is computed has a dramatic effect.  
To see this explicitly note that, at large $\nu\gg 1$, 
 $G_0\to \gamma^3 (\nu^3 - 3\nu)$ and 
 $\avg{x|\gam,\gam\nu}_{\rm ESP}\to \gamma\nu$.  
This makes 
\be
 f_{\rm ESP}(\nu) \to \frac{{\rm e}^{-\nu^2/2}}{\sqrt{2\pi}}\,(\nu^3-3\nu)\,
                          \frac{V\gamma^3}{V_\ast}
                \approx f_{\rm MS}(\nu)\,\frac{V\gamma^3\nu^3}{V_\ast}. 
\label{vfv-lim}
\ee
For $P(k)\propto k^n$, $\gamma^3V/V_* = [(n+3)/6]^{3/2}$ is just a 
constant independent of $m$, so
 $f_{\rm ESP}/f_{\rm MS}\to [\nu^2(n+3)/6]^{3/2}$; this grows rapidly 
at large $\nu$.  We will return to this shortly.  

\begin{figure}
 \centering
 \includegraphics[width=\hsize]{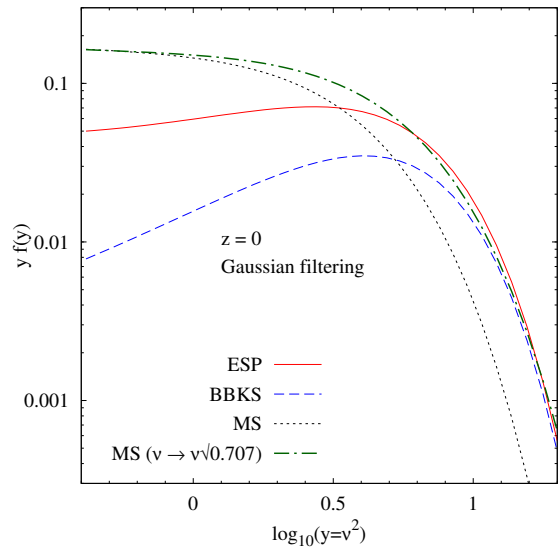}
 \caption{Mass functions with Gaussian filtering of a \Lam CDM
   spectrum in the three formalisms discussed in the text: 
   peaks (equation~\ref{vfv-bbks}, dashed blue), 
   excursion sets (equation~\ref{vfv-ms}, dotted black) and 
   ESP (equation~\ref{vfv-esp}, solid red). 
   The dot-dashed green curve shows \eqn{vfv-ms} with 
   $\nu\to\sqrt{0.707}\nu$; at large $\nu$, it is quite 
   similar to ESP. As discussed in the text, the dashed blue curve 
   for peaks is not well motivated and is only shown for comparison
   with the more appropriate ESP curve.
   }
 \label{esp-vfv}
\end{figure}

\fig{esp-vfv} compares these different models for halos 
using Gaussian filtering of the same \Lam CDM spectrum as before.
The curves show the results for BBKS (dashed blue), MS (dotted black)
and ESP (solid red).  The dashed blue curve is anecdotal, since, 
as we argued, it is not well motivated.  We have included it to 
illustrate that the difference between it and the more careful 
calculation (ESP) is small at large masses.  

The real interest in this plot is the fact that, at large masses, 
$f_{\rm MS}$ lies  about an order of magnitude below $f_{\rm ESP}$.  
Although we argued that this is expected (equation~\ref{vfv-lim}), 
it is interesting to consider this in view of our remarks in the 
Introduction about the discrepancy between the usual excursion set 
prediction and halo abundances in simulations.  We noted that to 
fit halo abundances it was common to scale the usual excursion set 
prediction (i.e. $f_{\rm MS}$) by setting $\nu \to a\nu$, with 
$a\sim \sqrt{0.707}$ (e.g. our Figure~\ref{vfv-th}).  
The dot-dashed green curve in \fig{esp-vfv} shows that setting 
$\nu\to\sqrt{0.707}\nu$ in \eqn{vfv-ms} brings it into 
remarkably better agreement with $f_{\rm ESP}$; values of $a$ 
between $0.7\sim0.8$ also do well.  
This strongly suggests that much of the discrepancy between the 
usual excursion set predictions and halo abundances in simulations 
can be attributed to inappropriate averaging in the excursion set 
approach.  We show in the next section that this has interesting 
consequences for the predicted spatial distribution of halos.

\subsection{Excursion set peaks with moving barriers}
\begin{figure}
 \centering
 \includegraphics[width=\hsize]{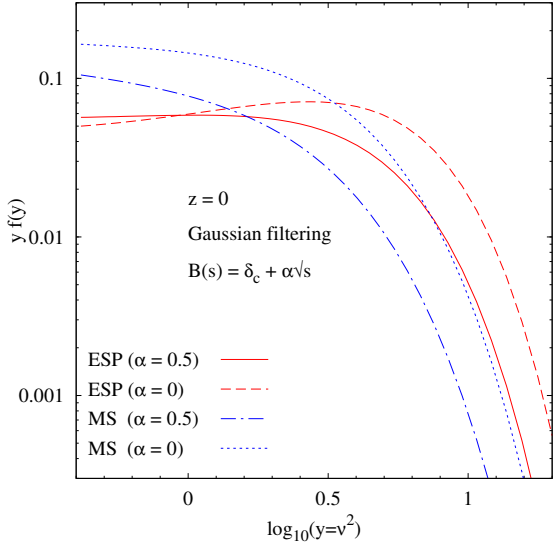}
 \caption{Mass functions with Gaussian filtering of a \Lam CDM
   spectrum for barrier shapes of the type $B(s)=\delc+\alpha\sqrt{s}$, 
   for excursion sets and ESP.
   The curves show ESP (red) for $\alpha=0.5$ (solid, 
   following from equation~\ref{Nesp-Bs}) and $\alpha=0$ (dashed, equation~\ref{vfv-esp}), 
   and the MS result (blue) for $\alpha=0.5$ (dot-dashed, their equation~5) 
   and $\alpha=0$ (dotted, equation~\ref{vfv-ms}).
   }
 \label{esp-moving}
\end{figure}

Before moving on to the study of predicted halo bias, it is worth 
noting that this approach makes it particularly easy to see how to 
incorporate the effects of a scale dependent $\delc$.  E.g., 
if we set $\delc\to B(s)$, then 
\be
 \Cal{N}_{\rm ESP}(\nu) = 
 \frac{1}{\gam\nu}\int_{2\gamma\sigma_0 B'}^\infty\der x\,(x-2\gamma\sigma_0 B')\,
      \Cal{N}_{\rm pk}(B/\sigma_0,x).
\label{Nesp-Bs}
\ee
In models of triaxial (rather than spherical) collapse, it is a 
good approximation to set $B(s)\approx \delc + \alpha \sqrt{s}$ 
\citep{smt01}, making 
 $B/\sigma_0 = \nu + \alpha$ and $2\gamma\sigma_0 B' = \gamma\alpha$.  
Figure~\ref{esp-moving} compares this model with $\alpha= 0.5$ 
(see Moreno et al. 2009 for why this value is interesting) with the 
case in which $\alpha=0$ ($B = \delc$ is constant).  This shows that 
while the shape of $f(\nu)$ is indeed sensitive to the physics of 
collapse, this sensitivity can only be used as a diagnostic if one 
is confident that the statistical prediction has been based on the 
correct ensemble average.

\section{Conditional mass functions and halo bias}
\label{mf-cond}
MS noted that it was straightforward to extend their analysis to 
make models of the mass fraction in halos of mass $m$ (corresponding to variance $s=\sig_0^2(R)$ with $R\propto m^{1/3}$) which are constrained to 
lie within regions of some specified size $R_0$ (corresponding to variance $S_0=\sig_0^2(R_0)$) and overdensity 
$\delta_0$.  An explicit expression for the conditional mass 
fraction, which is associated with the unconditional one in 
\eqn{vfv-ms} has recently been provided by \citet[MPS]{mps12}.  

It is instructive to rewrite their result in the notation of BBKS, 
who provided a similar analysis for peaks.  
Using the dictionary given in Appendix~\ref{bbkstomps}, 
equation~(28) of MPS can be written as 
\begin{align}
 \der\nu f_{\rm MPS}(\nu|\delo,\So) &=
  \der\nu_p\frac{{\rm e}^{-\nu_p^2/2}}{\sqrt{2\pi}}
  \frac{\avg{x|\tilde\gam,\tilde\gam\tilde\nu}_{\rm MS}}{\gam\nu}\,, 
\label{vfvcond-mps}
\end{align}
with $\avg{x|\gam,x_\ast}_{\rm MS}$ defined in \eqn{avgx-ms}, and the BBKS quantities $\{\nu_p,\tilde\gam,\tilde\nu\}$ given in \eqns{nup},~\eqref{gamtil} and~\eqref{gamtilnutil}.

Exactly the same logic as before, when applied to the BBKS expression for 
peaks conditioned on having $\delta_0$ on scale $S_0$ (equation~E11 of BBKS), 
leads to 
\begin{align}
 \der\nu f_{\rm ESP}(\nu|\delo,\So) &=
  \der\nu_p\frac{{\rm e}^{-\nu_p^2/2}}{\sqrt{2\pi}}\frac{V}{V_\ast}
   G_0(\tilde\gam,\tilde\gam\tilde\nu) \notag\\
&\ph{e^{-\nu^2/2}\sqrt{2\pi}}
\times \frac{\avg{x|\tilde\gam,\tilde\gam\tilde\nu}_{\rm ESP}}{\gam\nu}\,, 
\label{vfvcond-esp}
\end{align}
with $\avg{x|\gam,x_\ast}_{\rm ESP}$ defined in \eqn{avgx-esp}.

\subsection{Halo bias}
\label{bias}
MPS argued that a useful way of defining bias coefficients for a
Gaussian field is to cross-correlate the density of the biased tracers
with Hermite polynomials in the matter overdensity. The former is
defined as the ratio of the conditional and unconditional mass
fractions 
\be
 \avg{\rhoh|\delo} = f(\nu|\delo,\So)/f(\nu),
\ee
and MPS showed that applying this prescription to the excursion sets 
result (equations~\ref{vfvcond-mps} and ~\ref{vfv-ms}) leads to 
closed-form expressions for the bias coefficients:  
\begin{align}
b_n &\equiv \So^{-n/2}\avg{\rho_hH_n(\delo/\sqrt{\So})}\notag\\
&= \So^{-n/2}\int_{-\infty}^{\infty}\der\delo p_{\rm G}(\delo;\So)
\avg{\rhoh|\delo} H_n(\delo/\sqrt{\So})\,, 
\label{bn-def}
\end{align}
where $H_n(x)={\rm e}^{x^2/2}(-d/dx)^n {\rm e}^{-x^2/2}$ are the
``probabilist's'' Hermite polynomials. 

MPS also showed that these $b_n$ have the structure 
\be
b_n =\left(\frac{\Sc}{\So}\right)^n \sum_{r=0}^n \binom{n}{r}
b_{nr}\epc^r\,,  
\label{bn-expand}
\ee
where \Sc\ and \epc\ are given
in \eqn{Scepc}, and that the scale-independent (but mass-dependent)
$b_{nr}$ could be naturally interpreted as bias coefficients in
\emph{Fourier} space, with connections to the work of \citet{as88} 
and \citet{mat11}. They also showed that, at least for $n=1$,$2$, the $b_{nr}$
satisfy some remarkable linear relations between each other: for fixed
$n$, all the $b_{kr}$ with $1\leq r\leq k\leq n$ can be written as
linear combinations of $b_{k0}$, $1\leq k \leq n$. More surprisingly,
they showed that the peaks bias parameters at linear and quadratic
order derived by \citet{dcss10} \emph{also} satisfied exactly the same
linear relations between coefficients, although the form of the
coefficients themselves was different. 

Our results above allow us to generalise the connection between peaks 
and excursion sets bias to \emph{all} orders. Notice that the quantity
$\avg{\rhoh|\delo} = f(\nu|\delo,\So)/f(\nu)$ is given by
\be
\avg{\rhoh|\delo,\So} = \frac{\der\nu_p\,{\rm
    e}^{-\nu_p^2/2}/\sqrt{2\pi}}{\der\nu\,{\rm
    e}^{-\nu^2/2}/\sqrt{2\pi}} \,
\frac{G_J(\tilde\gam,\tilde\gam\tilde\nu)}{G_J(\gam,\gam\nu)}\,. 
\label{rhohgivend0}
\ee
This expression applies to all three formalisms with appropriate 
choices for $F(x)$ and $J$: for excursion sets $F(x)=1$, $J=1$, 
while for peaks and ESP we use \eqn{eqn-bbks-Fx} for $F(x)$, 
with $J=0$ for peaks and $J=1$ for ESP.
As a result, the only real difference between $\avg{\rhoh|\delo}$
defined for peaks, excursion sets or the ESP extension is in the
choice of the curvature function $F(x)$.  
This function is independent of \delo\ and simply goes for a
ride in the series expansion that defines the bias coefficients. More
precisely, the MPS result, that the $b_n$ are Taylor coefficients of
the expansion of $\avg{\rhoh|\delo,\tilde\cb=0}$ in powers of
\delo\ (where $\tilde\cb$ is the matrix given in
equation~\ref{ctil-def}), relied only on the properties of the
Gaussian $\sim{\rm e}^{-\nu_p^2/2}p_{\rm
  G}(x-\tilde\gam\tilde\nu;1-\tilde\gamma^2)$ and not on the fact that
they were analysing the special case $F(x)=1$. This result therefore
applies equally well to peaks theory and its extension.  

This allows us to 
generalise the results in Appendix A of MPS trivially.  
We show how to do this in Appendix~\ref{mpstogenericbias}, and we find 
\be
\delc^nb_n = \left(\frac{\Sc}{\So}\right)^n
\sum_{k=0}^n\binom{n}{k}(1-\epc)^k\lam_k\mu_{n-k}\,, 
\label{bn-generic}
\ee
with the $\nu$-dependent quantities $\mu_k$ and $\lam_k$ defined as  
\be
\mu_k\equiv \nu^kH_k(\nu) ~~;~~ \lam_k\equiv
(-\Gam\nu)^k\avg{H_k(y-\Gam\nu)}_F\,,  
\label{muklamk}
\ee
where $\Gam^2\equiv\gam^2/(1-\gam^2)$ and the $F$-averaged Hermite
polynomial is 
\begin{align}
&\avg{H_k(y-y_\ast)}_F \notag\\
&\ph{Hk}
\equiv \frac{\int_0^\infty\der y\,y^J\,F(y\gam/\Gam)p_{\rm G}(y-\Gam\nu;1)
  H_k(y-y_\ast)}{\int_0^\infty\der y\,y^J\,F(y\gam/\Gam)p_{\rm  G}(y-\Gam\nu;1)}\,.
\label{gavg}
\end{align}
It is straightforward to check that setting $F(x)=1$, $J=1$ recovers the
results of MPS, while setting $J=0$ and using \eqn{eqn-bbks-Fx} with $n=1$,$2$ 
recovers those of \citet{dcss10}. More interestingly, \eqn{bn-generic}
can be rearranged to write \eqn{bn-expand} with
\be
\delc^nb_{nr} = (-1)^r\sum_{k=r}^n\binom{n-r}{k-r}\mu_{n-k}\lam_k\,.
\label{bnr-generic}
\ee
The $\mu_k$ are independent of the function $F(x)$, so that it is
useful to set $r=0$ and re-express the $F$-dependent $\lam_k$ in terms
of the peak-background split parameters $b_{n0}$. Using
$\mu_0=1=\lam_0$ we can write $\delc^nb_{n0}-\mu_n =
\sum_{k=1}^nD_{nk}\lam_k$, where $D_{nk}\equiv\binom{n}{k}\mu_{n-k}$
is an invertible lower-triangular matrix with diagonal elements unity,
and we find 
\begin{align}
\lam_1 &= \delc b_{10}-\mu_1\,,\notag\\
\lam_n &= \delc^nb_{n0} - \mu_n - \sum_{k=1}^{n-1}\binom{n}{k}\mu_{n-k}\lam_k ~~,~ n>1\,.
\label{lamk-inverted}
\end{align}
This explicitly demonstrates how all the $b_{kr}$ for $r\leq k\leq n$
can be expressed in terms of the $b_{k0}$ with $k\leq n$, for
arbitrary $n$, thus generalising the MPS result for
$n=1$,$2$.  This algebraic structure is clearly
independent of the choice of $F(x)$ and $J$ and also holds for peaks
theory and ESP. Furthermore, the expressions
above are extremely simple ways of calculating bias parameters at
\emph{any} order (compared, e.g., to the painstaking calculation in 
\citealt{dcss10} for $n\leq2$).

\begin{figure}
 \centering
 \includegraphics[width=\hsize]{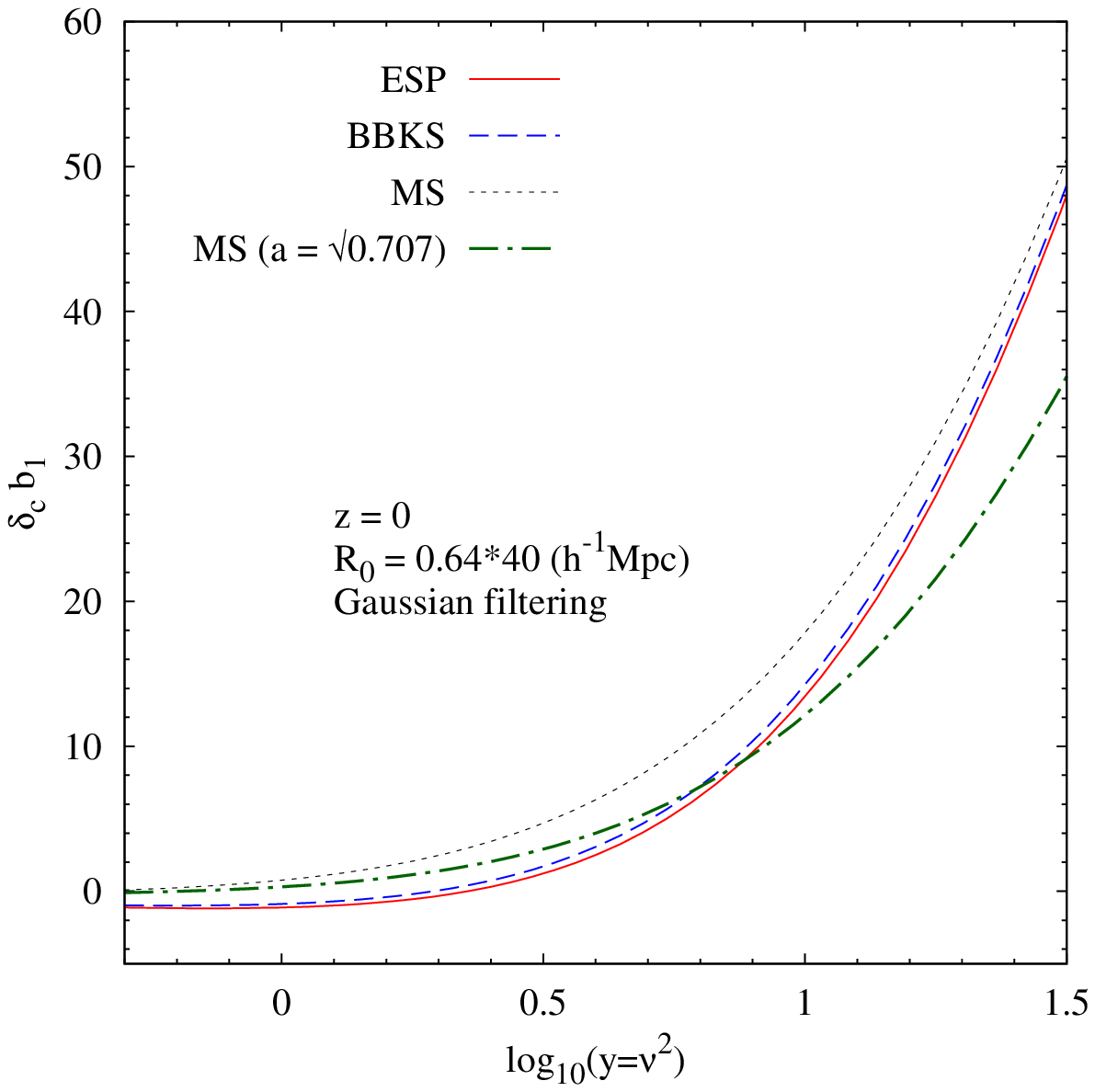}
 \caption{Linear bias predicted by peaks, excursion sets and ESP for a
   Gaussian filtered \Lam CDM spectrum, coded as in \fig{esp-vfv}. 
   The cross-correlation in \eqn{bn-def} was defined on a Lagrangian
   scale $R_0=0.64\times40 h^{-1}$Mpc (at which the Gaussian filter
   encloses the same mass as the TopHat filter at $40 h^{-1}$Mpc).
   Notice that whereas the replacement
   $\nu\to\sqrt{0.707}\nu$ in the MS mass function improved its
   agreement with the peaks and ESP mass functions at large masses,
   the same replacement under-predicts the linear bias at large
   masses. Interestingly, this is qualitatively similar to what is
   seen in $N$-body simulations when comparing TopHat filtered mass
   functions and their associated linear bias relations. See text for
   a discussion.}
 \label{esp-bias}
\end{figure}

\fig{esp-bias} shows the predicted linear bias for peaks, excursion
sets and ESP for Gaussian filtering of the same \Lam CDM spectrum and
with the same colour-coding as in \fig{esp-vfv}, with the
cross-correlation in \eqn{bn-def} defined on a Lagrangian scale
$R_0=0.64\times40 h^{-1}$Mpc (at which the Gaussian filter encloses
the same mass as the TopHat filter at $40 h^{-1}$Mpc). All three
formalisms predict that $\delc b_1$ approaches $\sim(\Sc/\So)\nu^2$ at
large masses. For this reason, the modified excursion set result with
$\nu\to a\nu$ (dot-dashed green) predicts a linear bias that
is below the one for peaks and ESP at large masses. This is
interesting because it is qualitatively the same as what has been
recently found in $N$-body simulations: at large masses, an 
analytical mass function such as the one of \citet{st99} with 
$a<1$ chosen to match the mass function of an $N$-body simulation 
\emph{under-predicts} the linear halo bias measured in the same 
simulation \citep{mss10, t+10}. This suggests that an analysis based
on peaks theory such as the one presented here is on the right track
towards obtaining an accurate description of both the mass function
and the halo bias from first principles.

\section{Discussion}
We showed that, especially if one accounts for correlations between steps, the standard, spherical-collapse based excursion set prediction for halo mass fractions (equation~\ref{vfv-ms}) vastly underestimates what is measured in simulations.  However, rescaling the spherical collapse motivated self-similar scaling variable $\nu\to\sqrt{0.7}\nu$ results in much better agreement (Figure~\ref{vfv-th}).  We noted that this agreement should not be used to argue that halos formed from a spherical collapse -- at least, not until the reason for the adhoc rescaling of $\nu$ is understood.  

We then argued that the rescaling was related to a flaw in the usual formulation of the excursion set approach \citep{bcek91}, in which one replaces an average over spatial positions in one realization of the field with an ergodic average over many independent realizations of the field.  Although we are not the first to have noted this problem, much previous work has attempted to rectify this by accounting for spatial correlations between walks.  However, measurements in simulations showing that halos form around special positions in the initial fluctuation field \citep{smt01} suggest that it may be more productive to instead modify the ensemble over which one computes statistical averages.  

We then used peaks theory to illustrate this point, by showing how to incorporate the peaks constraint into the excursion set formalism.  Specifically, peaks correspond to regions around which the curvature of the local density is modified (BBKS), and this, we argued, modifies the excursion set prediction from \eqn{vfv-ms} to \eqn{vfv-esp}.  In fact, the fundamental role played by the curvature distribution $F(x)$ (equation~\ref{eqn-bbks-Fx}) in our analysis suggests that to build an accurate model of halo abundances, all one needs is a good model for the initial profile shapes from which halos form.  For example, one might combine measurements of the density run around virialized halos with infall models to infer what the initial overdensity profiles must have been; having found them, one could use them instead of $F(x)$ in \eqn{vfv-esp} and so predict the halo mass fraction $f(\nu)$.  This is in progress.  

Although our analysis has gone some way towards addressing the real cloud in cloud problem (correlated steps and correlated walks), there is more that can be done in this direction.  This is because our analysis is fundamentally about taking `one small step' beyond that on which the object was defined; therefore, it does not correctly account for small objects which are embedded in much more massive objects (i.e., when the smoothing scales are rather different).  Some of the nicest work in this direction is in a series of papers by \cite[and references therein]{m+98}; we are in the process of incorporating their work into our analysis.  

Our formulation of peaks in the excursion set language made it particularly easy to see how to build an excursion set model for peaks even when the question of which peaks are interesting depends on smoothing scale -- the analogue of moving barriers in the excursion set approach (equation~\ref{Nesp-Bs} and~Figure~\ref{esp-moving}).  This may prove necessary if one wishes to incorporate the effects of the stochasticity associated with non-spherical collapse into the excursion set peaks predictions.  

The similarity in formulation also allowed a simple description of how peak abundances are modified if the large scale density field is constrained to be different in some way (equation~\ref{vfvcond-esp}).  In turn, this allowed a simple generalization of earlier results on peak and halo bias to all orders (equations~\ref{bn-generic}--\ref{lamk-inverted}).  In particular, we showed that excursion set peak bias is most easily understood in Fourier space, where it is $k$-dependent even at the linear level.  

Although we concentrated on an excursion set analysis of peaks, the MS `one-step' argument should apply to other traditionally single-scale analyses of cosmological datasets.  For example, since the argument is not restricted to three dimensional fields, it can be applied to interpret the CMB temperature distribution, which is a two dimensional (nearly if not exactly Gaussian) random field.  The number density and clustering of `hotspots', as a function of spot temperature, has been used as a diagnostic of the Gaussianity of this field \citep{be87, hs99}.  But since some hot spots will be local maxima on larger smoothing scales as well, it is of interest to describe how the distribution of sizes (and the clustering) of regions which lie above some threshold temperature depends on the value of threshold.  Clearly, the analysis presented here can be applied to that problem directly.  In three dimensions, perhaps the most interesting connection and application is to the series of recent papers on the `skeleton' of the cosmic web \citep{skeleton}.  This is the subject of ongoing work, where we hope to make a connection to the multi-scale analyses of \citet*{ac10}.

Although essentially all the analysis in this paper used Gaussian smoothing filters (section~\ref{notation} discussed why, in the present context, they simplify the analysis substantially), we do not think they are otherwise special, so we are in the process of extending our results to include tophat smoothing filters.  Since fitting functions for halo counts in simulations use tophat filtering (for the conversion between $\sigma$ and halo mass) exclusively, until our analysis does the same, a direct comparison with measurements of halo mass functions in simulations is premature.  
%(There is a sense in which the Gaussian filters are special -- some of the integrals which enter peaks theory are ill-defined for tophat filters.  This is why we believe the extension beyond Gaussian filters is best presented separately.)

This is particularly interesting in view of the fact that the linear bias factor in our excursion set peaks model is close to the usual excursion set predictions associated with rescaled $\delc$ at small masses, but with no rescaling of $\delc$ at high masses (Figure~\ref{esp-bias}).  This last is in qualititive agreement with measurements of halo bias in simulations.  We believe that matching the enhanced abundance and bias at large masses (\figs{esp-vfv} and~\ref{esp-bias}), without having to rescale the parameter which is associated with the physics of halo formation, are nontrivial and encouraging successes.  

\section*{Acknowledgements}
We thank V. Desjacques for useful discussions.
This work is supported in part by NSF-0908241 and NASA NNX11A125G.
RKS thanks the GEPI and LUTH groups at Meudon Observatory for hospitality
during the summer of 2012.

\appendix
\section{Technical details}
\label{app-details}
In this Appendix we collect technical details of various results
quoted in the text.

\subsection{From BBKS to MPS}
\label{bbkstomps}
For Gaussian filters, the dictionary for converting between $(\tilde\gam,\tilde\nu,\nu_p)$
and the quantities $(\Sc,\epc,\dcr,Q,\dprbar,\sigdpr)$ defined by MPS (they denoted $v\equiv \der\del/\der s$ as $\del^\prime$) is  
\begin{align}
\nu_p &= \frac{\dcr}{\sqrt{sQ}} \,,
\label{nup}\\
1-\tilde\gamma^2 &= 4s\gam^2\sigdprsq =
\frac{\sigdprsq}{\avg{v^2}} = \frac{{\rm
    Var}(v|\nu,\delo)}{{\rm  Var}(v)}\,,
\label{gamtil}\\
\tilde\gam\tilde\nu &= 2\gam\sqrt{s}\dprbar =
\frac{\avg{v|\nu,\delo}}{\sqrt{{\rm Var}(v)}}\,, 
\label{gamtilnutil}
\end{align}
where 
\begin{align}
& \dcr \equiv \delc - \delo\frac{\Sc}{\So} ~~;~~ Q\equiv
  1-\left(\frac{\Sc}{\So}\right)^2\frac{\So}{s}\,,\notag\\ 
& \dprbar \equiv \avg{\!v|\nu,\delo\!} 
 = \frac1{2sQ}\!\left[ \dcr +
 \epc\frac{\Sc}{\So}\bigg(\delo -
 \delc\frac{\Sc}{\So}\frac{\So}{s}\bigg)\right],\notag\\ 
& \sigdprsq \equiv {\rm Var}(v|\nu,\delo)
 = \frac1{4\Gamma^2s}\! \left[1-\frac{\Gamma^2\So}{Qs} 
 \frac{\Sc^2 (1-\epc)^2}{\So^2} \right]\,,
\label{dcrQdprsigbar}
\end{align}
with 
\begin{align}
\Sc &= \avg{\del\delo} ~;~ \epc = \frac{2s}{\Sc}\avg{v\delo} = 2\frac{\der\ln\Sc}{\der\ln s}\,. 
\label{Scepc}
\end{align}
Related to these is the matrix $\tilde\cb$ discussed in
section~\ref{bias}, which is the quantity
one must subtract from the unconditional covariance matrix of the
variables $(\del,v)$ to obtain the covariance matrix of the
conditional Gaussian $p(\del,v|\delo)$. This follows from
equation~(14) of MPS:  
\be
  \tilde\cb = \frac{\Sc^2}{s\So}\!
  \left[\!
  \begin{array}{cc} s & \epc/2 \\ \epc/2 & \epc^2/4s \end{array}
  \!\right]\,.
\label{ctil-def}
\ee

\subsection{Generalising the MPS results for bias}
\label{mpstogenericbias}
For the reasons mentioned in section~\ref{bias}, it is straightforward
to generalise the results of MPS for halo bias to include peaks theory
and its extension discussed in this work. To do this, we note that the
results in their Appendices~A.2 and~A.3 only depend on the form of the
conditional Gaussian distribution $p(\nu,x|\delo)$ (they
work with $p(\del,v|\delo)$) and not
on the fact that their integrand of $x$ used $F(x)=1$.
All that is needed then is to extend the
results of their Appendix~A.4 to include an arbitrary function $F(x)$
and value $J$
in calculating the quantity $\avg{\rhoh|\delo,\tilde\cb=0}$. 
With $\Gam^2\equiv\gam^2/(1-\gam^2)$, equation~(A8) of MPS can be 
generalised to obtain  
\begin{align}
&\avg{\rhoh|\delo,\tilde\cb=0} \notag\\
&\ph{rhoh}
= {\rm
  e}^{\frac12\nu^2-\frac12\nu^2(1-\bar\del_0\Sc/\So)^2} \notag\\ 
&\ph{rhoh|\delo}
\times
\frac{\int_0^\infty\der y\,y^J\,F(y\gam/\Gam)p_{\rm
    G}(y-\Gam\nu+\bar\del_0\nu_1;1)}{\int_0^\infty\der
  y\,y^J\,F(y\gam/\Gam)p_{\rm  G}(y-\Gam\nu;1)}\,,
\label{rhoh-ctil=0}
\end{align}
where we used $y=x\Gam/\gam$ and followed MPS in defining
$\bar\del_0\equiv\delo/\delc$ and
$\nu_1\equiv\Gam\nu(\Sc/\So)(1-\epc)$. 

This can also be seen more directly as follows.
As MPS discussed, the condition $\tilde\cb=0$ corresponds to the
assignments  
\begin{align}
&Q\to1 ~;~~ \sigdpr\to(2\Gam\sqrt{s})^{-1} ~;~~
  \dprbar/\sigdpr\to(\Gam\nu-\bar\del_0\nu_1)\,, 
\label{ctil=0}
\end{align}
or
\begin{align}
\nu_p&\to\nu(1-\bar\del_0(\Sc/\So))\,,\notag\\ 
1-\tilde\gam^2 &\to 1-\gam^2\,,\notag\\
\tilde\gam\tilde\nu &\to \gam\nu(1-\bar\del_0(\Sc/\So)(1-\epc))\,.
\label{ctil=0-bbks}
\end{align}
Making these replacements in \eqn{rhohgivend0} gives
\eqn{rhoh-ctil=0}. Taylor expanding this expression gives the bias
coefficients as $\avg{\rhoh|\delo,\tilde\cb=0} =
\sum_{n=0}^\infty\bar\del_0^n(\delc^nb_n)/n!$. The expansions of both
Gaussians in \eqn{rhoh-ctil=0} involve Hermite polynomials, and lead
to \eqn{bn-generic}.

\label{lastpage}

\end{document}